\documentclass[12pt]{article}
  \newcommand{\be}[3]{\begin{equation}  \label{#1#2#3}}
  \newcommand{\bib}[3]{\bibitem{#1#2#3}}
\newcommand{\ee}{ \end{equation}}
\newcommand{\ba}{\begin{array}}
\newcommand{\ea}{\end{array}}
\newcommand{\p}{\partial}
\renewcommand{\arraystretch}{1.7}
\let\LARGE=\Large
\let\Large=\large

 \def\unit{\hbox to 3.3pt{\hskip1.3pt \vrule height 7pt width .4pt \hskip.7pt
\vrule height 7.85pt width .4pt \kern-2.4pt
\hrulefill \kern-3pt
\raise 4pt\hbox{\char'40}}}

\begin{document}


\thispagestyle{empty}
\rightline{HUB-EP-98/38}
\rightline{DAMTP-1998-73}
\rightline{hep-th/9806195}

\vspace{1.5truecm}

\centerline{\bf \LARGE ${\bf AdS_3}$ Gravity and Conformal Field
 Theories}
\vspace{2truecm}

\centerline{
  {\bf Klaus Behrndt}$^{a}$ , {\bf Ilka Brunner}$^{a}$
  and
  {\bf Ingo Gaida}$^{b}$
 \footnote{e-mail: behrndt@physik.hu-berlin.de,
                   brunner@physik.hu-berlin.de,
                   I.W.Gaida@damtp.cam.ac.uk,
Research supported by Deutsche Forschungsgemeinschaft (DFG).
  }
 }
\vspace{.5truecm}

\centerline{\em $~^a$Humboldt University Berlin}
\centerline{\em Invalidenstrasse 110, 10115 Berlin, Germany}
\vspace{.3truecm}
\centerline{\em $^b$ DAMTP, University of Cambridge}
\centerline{\em Silver Street, Cambridge CB3 9EW, UK }



\vspace{.5truecm}

\begin{abstract}
We present a detailed analysis of $AdS_3$ gravity, the
BTZ black hole and the associated conformal field theories
(CFTs).
In particular we focus on the non-extreme
six-dimensional string solution
with background metric $AdS_3 \times S^3$ near the horizon.
In addition we introduce momentum modes along the string,
corresponding to a BTZ black hole, and a Taub-NUT soliton
in the transverse Euclidean space. We show that the $AdS_3$ space-time of
this configuration has the spatial geometry of an annulus with
a Liouville model at the outer boundary and a two-dimensional
black hole at the inner boundary. These CFTs
provide the dynamical degrees of freedom of the three-dimensional
effective model
and, together with the CFT corresponding to $S^3$, provide
a statistical interpretation of the
corresponding Bekenstein-Hawking entropy.
We test the proposed exact black hole entropy, which should hold
to all orders in $\alpha^\prime$,
by an independent field theoretical analysis including
higher-order curvature corrections. We find consistent results that
yield a renormalization of the classical parameters, only.
In addition we find a logarithmic subleading black hole entropy
coming from gravitational fivebrane instantons in a special limit in
moduli space.

\vspace{0,5cm}

\end{abstract}


\newpage


\section{Introduction}

In the last years (intersecting) D-branes provided a complete new
point of view in black hole physics and in gauge
theories living on the D-brane world volume. They may provide
a universal link between Yang-Mills theory and gravity. As solutions of
the corresponding supergravity equations of motion,
branes are typically singular
indicating a strong interaction between the world volume theory and
the bulk gravity. However, there is a certain subclass of branes
where both theories may decouple in certain limits. These non-singular
branes are scalar-free and space-time becomes a product space
near the horizon, i.e. it factorizes
into an anti-de Sitter ($AdS$) space-time,
a spherical space and a flat Euclidean space
($AdS_{p} \times S_{q} \times E_r$).
The diffeomorphism group of
$AdS$ space-time manifests itself as a conformal group living on the boundary
of $AdS$. Thus it is natural to relate physics
on the brane with a conformal field theory (CFT) living on the
boundary of $AdS$. In fact, in the context of the supermembrane
this idea has been proposed already ten years ago \cite{132}.
\\
In a quasi-classical approximation ($\alpha' \rightarrow 0$ and for
large $N$) concrete suggestions have been made also for other
brane configurations \cite{110}. A great deal of attention
received the cases of odd $AdS$ space-times; for example the CFT on the
boundary of $AdS_7$ is expected to be dual to a non-critical string
theory describing the worldvolume of the M5-brane \cite{100}. The
boundary CFT of $AdS_5$ is believed to be dual to 4-d super Yang-Mills
theory describing the D3-brane world volume theory \cite{120}, \cite{122}.
Moreover the 2-d CFT on the boundary of $AdS_3$ should be dual to a 2-d
$\sigma$-model describing the world volume theory of the D1-brane
\cite{400}.  As stressed in \cite{130} these odd $AdS$
space-times are
symmetric on both sides of the horizon. There is no singularity beyond
the horizon and one may address the question, what a horizon means in
the related CFT.
\\
For the $AdS_7$ and $AdS_5$ examples it is important
to consider special limits where
one can trust these solutions ($\alpha' \rightarrow 0$ and
large $N$). Although, keeping enough supersymmetries, these geometries
may correspond to exact backgrounds, even at the quantum level
\cite{124}. On the other hand, for $AdS_3$ the conformal
field theory on the boundary becomes infinite dimensional and, as a
consequence of this symmetry enhancement, we do not need to consider
$\alpha'\rightarrow 0$ and/or large $N$.  These models are
well-defined for all $N$ or $\alpha'$ - even without supersymmetry.
Furthermore, because $AdS_3$ is the near-horizon geometry of strings
in six or five dimensions, the CFT will provide a microscopic picture of
the black hole entropy including $\alpha'$ corrections. Thus, there are
especially two motivations to consider $AdS_3$ configurations:

(i) To obtain finite $N$ results. Especially one should address the
question of phase transitions which may spoil the
limit from large $N$ to finite $N$ \cite{126}.

(ii) To determine $\alpha'$ corrections. Applied to the black hole entropy
these are corrections coming, for instance,
from higher-order curvature corrections.

As we will see below, the exact results will not only fine-tune the
lowest order results, instead qualitative new features will appear, too.
\\
Anti-de Sitter gravity in three dimensions,
i.e.\ Einstein-Hilbert action plus negative
cosmological constant, has a long history. Key
observations are: (i) The $SL(2, {\bf R})_R  \times SL(2, {\bf
R})_L$ conformal group on the boundary is enhanced to an infinite
dimensional Virasoro algebra \cite{150}; (ii) there exist a
reformulation
as a topological Chern-Simons gauge theory \cite{Achucarro} with
no bulk degrees of freedom.  However, if the manifold has non-trivial
boundaries, gauge degrees of freedom become dynamical at the boundary.
It follows \cite{440} that the boundary field theory is
given by Wess-Zumino-Witten (WZW) models.
\\
A concrete realization has been discussed by Carlip \cite{424} (and
later in \cite{420}), who showed that on the asymptotic boundary a
Liouville model is realized. Soon afterwards Banados, Teitelboim and
Zanelli (BTZ) found a black hole solution of  $AdS_3$ gravity
\cite{170} (see also \cite{172}).
Since then many aspects of these black holes have
been discussed. The entropy and the microscopic understanding in
terms of the boundary CFT in \cite{174} - \cite{630}, in \cite{Welch}
the CFT has been analysed and it has been shown
that the BTZ black hole is $T$-dual to
a black string \cite{540,500}. Apart from the fact that the BTZ black
hole is a solution of $AdS_3$ gravity, it is also a good example for
so-called topological black holes that can be obtained by discrete
identifications in anti-de Sitter space-times \cite{430}.
\\
In this article we focus on the discussion of subleading terms,
i.e.\ $\alpha'$ or finite $N$ corrections to the near-horizon geometry of
a non-extreme string in six dimensions, which
corresponds in ten dimensions to a string living inside a 5-brane.
In addition to these two branes we include a Taub-NUT soliton
and add momentum modes along the string. In the near-horizon
region, the Taub-NUT soliton yields an orbifolding of the $S_3$ and
the waves produces a BTZ black hole in the $AdS_3$ part (section 2).
At the same time the black hole provides an additional
boundary in $AdS_3$, i.e.\ the spatial three-dimensional
geometry becomes an annulus.  On both boundaries live {\em different}
CFTs (section 4): on the asymptotic boundary it is a Liouville model
(as expected) and on the horizon it is a 2-d black hole (see also
\cite{520}).  In a previous paper \cite{412} we discussed already the
entropy and subleading contribution to the central charge and obtained
$c = 6 k + \beta + {\gamma \over k}$, where $k$ is the Chern-Simons
level and $\beta$ and $\gamma$ are some numbers. Here we work
out the complete CFT and interprete the subleading contribution as
$\alpha'$ corrections, which fit with higher-order curvature corrections to
the Bekenstein-Hawking entropy of 4-d black holes (section 5).
In addition we review in section 3 the different $AdS_3$
parametrizations and the discrete identifications, which yield the
BTZ black hole.

\section{The near-horizon region of strings}

As a ``master model'' we can consider the 4-charge configuration of
the $NS$-sector including a fundamental string, a 5-brane, a wave and
a Taub-NUT soliton. This configuration is part of all string
models. In particular it is $S$-dual to the D1-D5 system
and it contains enough charges to address the question of the exact
entropy of four dimensional black hole solutions carrying 4 charges.
The corresponding metric can be obtained from the
the non-extreme string background metric \cite{423}
\be098
ds^2 = {1 \over H_1} \Big( -(1-{\mu \over r} )\, dt^2 + dy^2 \Big) + H_2
        \left( {1 \over H_3}
        (dx_4 + \vec{V} d\vec{x})^2 + H_3 ({dr^2 \over 1 - {\mu\over r}} +
        r^2 d\Omega_2) \right)
        + ds^2_{int}
\ee
and, after a finite boost along the string direction
\be099
dt \ \rightarrow \ \cosh\beta \, dt + \sinh\beta \, dy \qquad , \qquad
dy \ \rightarrow \ \sinh\beta \, dt + \cosh\beta \, dy,
\ee
one obtains
\be100
\ba{l}
ds^2 = {1 \over H_1} \Big( -dt^2 + dy^2 + {q_0 \tanh \beta \over r}
        (dy - \coth\beta dt)^2 \Big)  \\
\qquad + H_2 \left( {1 \over H_3}
        (dx_4 + \vec{V} d\vec{x})^2 + H_3 ({dr^2 \over 1 - {\mu\over r}} +
        r^2 d\Omega) \right)
        + ds^2_{int}\\
H =  d (1/H_1) \wedge dt \wedge dy + \;^{\star} (dH_2 \wedge dt \wedge dy)
\quad , \quad e^{-2\phi} = {H_1 \over H_2} \quad , \quad
dH_3 = \,^{\star} dV
\ea
\ee
with $q_0 \equiv \mu \cosh\beta \sinh\beta$.
In the extreme limit ($\beta \rightarrow \infty$)
one keeps $q_0$ fix, which becomes the wave
charge. Moreover, every harmonic function parametrizes one brane:\\[1mm]
$H_1= 1 + {q_1 \over r}$: the fundamental string \newline
$H_2= 1 + {p^2 \over r}$: (compactified) NS-5-brane \newline
$H_3= 1 + {p^3 \over r}$: KK-monopole (Taub-NUT space)\\[1mm]
and $ds^2_{int}$ is the 4-dimensional compact space (5-brane
worldvolume), which we will assume
to be trivial, e.g.\ it is given by a torus. In the following
we will omit this internal space. Thus, introducing polar
coordinates one obtains for the near-horizon geometry
\be130
\ba{rcl}
ds^2 &=& {r \over q_1} \left[ -dt^2 + dy^2 + {q_0 \tanh\beta \over r}
        (dy - \coth\beta dt)^2 \right]  + p^2 p^3 \left({dr^2 \over
        r(r-\mu)}\right)^2 + \\
&& {p^2 p^3} \left[ (d\zeta + (\pm 1-\cos\theta) \, d\phi)^2 +
         d\Omega_2  \right] \\
H = H_1 + H_2 &=& {1 \over q_1} \left(dr  \wedge dt \wedge dy\right) +
        p^2 p^3 \left( d\zeta  \wedge d\phi \wedge d\theta \right)
        \quad , \quad e^{-2\phi} = {q_1 \over p^2} \ .
\ea
\ee
with $x^4 \equiv p^3 \zeta$ and the ``$\pm $'' ambiguity indicates the
different choices for the north and south hemisphere.
It follows that near the horizon ($r=0$)
the six-dimensional space-time becomes
a product space of two three-dimensional subspaces
\be140
 M_{6} =  {AdS}_{3} \ \times \  S^{3}/Z_m \ .
\ee
The Euclidean space $S^3/Z_m$ is described by a $SU(2)/Z(m)$-WZW model
(see \cite{Lowe} and reference therein), a discrete subgroup is
projected out due to the KK-monopole ($\zeta \simeq \zeta + {4\pi
\over m}$). This model corresponds to an {\em exact} three dimensional
conformal field theory. Its level is related to the radius of
the $S_3$ and the central charge is given by
\be150
k \equiv k_{SU} = {p^2 p^3 \over \alpha'}\ , \ \ \ \
c_{SU} = {3k \over k + 2}  \ .
\ee
In the classical limit $k \rightarrow \infty$ (or $\alpha'
\rightarrow 0$) one obtains $c_{SU} =3$.  Later on, we will be interested
in the entropy of 4-dimensional black holes. Thus, we
dimensionally reduce over the NUT-direction.
This reduces the effective central charge by one.
Moreover, since we have a compact group
manifold $k$ has to be quantized (positive integer). The non-compact
three-dimensional space-time $AdS_3$, on the other hand, represents the
Banados-Teitelboim-Zanelli (BTZ) black hole \cite{170,172}
for appropriate values
of the charges with coordinates $x^\mu = (t,y,r)$.  Starting with
(\ref{130}) one can perform the following transformation
\be160
t \rightarrow \sqrt{ q_1 \over l} \; t \qquad , \quad
y \rightarrow \sqrt{q_1 \over l} \; y \qquad , \qquad
r \rightarrow  {r^2 \over l} - q_0 \tanh\beta
\ee
to obtain the  metric:
\be170
ds^2 = - e^{-2V(r)} \ dt^2 + e^{2V(r)} \ dr^2 + \Big({r\over l}\Big)^2 \
        \Big( dy -\frac{r_- r_+}{ r^2} \, dt  \Big)^2
\ee
with
\be180
e^{-2 V(r)} = {(r^2 - r_-^2)(r^2 - r_+^2) \over r^2 l^2}
        \  , \quad  r_{\pm}^2 = {l q_0 \over (\tanh\beta)^{\pm1}}
        \ ,\quad \mu = {r_+^2 - r_-^2 \over l}
        \  , \quad  l^2 = 4 p^2 p^3
\ee
The horizons of the BTZ black hole are located at $r=r_{\pm}$,
the mass and angular momentum are given by
$M = {r_+^2 + r_-^2 \over l^2}$,
$J= {r_+ r_- \over l^2}$, respectively.
The background metric solves the three-dimensional Einstein-Hilbert action
\be190
  S_{EH} = \frac{1}{2 \kappa_3^2} \int_{AdS_3} \ d^3 x \ \sqrt{-g} \
             (R \ - \ 2 \Lambda)
\ee
including a negative cosmological constant $\Lambda=-1/l^2$.
In the limit $q_0 \rightarrow 0$ one obtains the empty space solution
($AdS$ vacuum state) with metric
\be200
 ds_{\rm vac}^2 =  - \frac{r^2}{l^2} dt^2 +  \frac{l^2}{r^2} \ dr^2
          + \frac{r^2}{l^2} \  dy^2 \ .
\ee
In our ``master model'' we reach this vacuum solution if there are no
wave-modes along the six-dimensional string ($q_0 = 0$) and from the
point of view of the BTZ black hole it corresponds to the massless case
($M=J=0$). Note that the vacuum solution and the ``standard'' $AdS_3$
metric
\be210
 ds_{\rm AdS_3}^2 = - \left ( \frac{r^2}{l^2} + 1 \right ) \ dt^2
+ \left ( \frac{r^2}{l^2} + 1 \right )^{-1} \ dr^2
          + \frac{r^2}{l^2} \  dy^2.
\ee
can be locally mapped onto each other, but they are globally
inequivalent.

The six-dimensional string configuration is a solution of the action
\be220
 S_6 = \frac{1}{2 \kappa^2_6} \int_{{M}_{6}} d^6 x
 \sqrt{-G} e^{-2 \phi}
\left [
 R - \frac{1}{12} H_{\mu \nu \rho}  H^{\mu \nu \rho}
 + 4 (\partial \phi)^2
\right ]
\ee
with $ e^{-2 \phi} = H_1 / H_2$.
Near the horizon $r=0$ the six-dimensional
space-time $M_6$ becomes a product space and, therefore,
\be230
\ba{rcl}
 \lim_{r \rightarrow 0} R({M}_6) &=&  R({AdS}_{3}) \ + \  R({S}^{3})
\\
 \lim_{r \rightarrow 0} \sqrt{-G({M}_6)} &=& \sqrt{-g({AdS}_{3})} \
                                              \sqrt{g({S}^{3})}.
\ea
\ee
Thus, using
\be340
\ba{rcl}
\int_{{S}^{3}} d^3 x \ \sqrt{g({S}^{3})} \  R({S}^{3}) &=& V({S}^{3}) \
        \frac{4}{l^2}
\\
\int_{{S}^{3}} d^3 x \ \sqrt{g({S}^{3})}  &=&  V({S}^{3})
\ea
\ee
one obtains near the horizon the following three-dimensional action
\be350
S_3 = e^{-2 \phi_h} \frac{V}{2 \kappa^2_6} \int_{{AdS}_{3}} d^3 x
         \sqrt{-g} \left [ R - \frac{1}{12} H_{\mu \nu \rho}
        H^{\mu \nu \rho}+ \frac{4}{l^2}
\right ].
\ee
Here $e^{2\phi_h}$ denotes the dilaton at the horizon $r=0$, which is
a constant.
The solution of the equations of motion is given by \cite{Welch}
\be370
H_{\mu\nu\rho}  =  \frac{2}{l} \ \epsilon_{\mu\nu\rho} , \hspace{2cm}
        R_{\mu\nu} = - \frac{2}{l^2} \ g_{\mu\nu}.
\ee
Since the two-form field strength is constant, one can use its equation
of motion to obtain from (\ref{350}) the Einstein-Hilbert action
in $AdS_3$:
\be390
S_{EH} =   \frac{1}{2 \kappa^2_3} \int_{{AdS}_{3}} d^3 x
 \sqrt{-g} \ \left ( R - 2 \Lambda \right ),
\hspace{1cm} \kappa^2_3 = \frac{\kappa^2_6}{V(S^{3})} e^{2\phi_h}.
\ee
Note that a 3-form field strength in 3 dimension is dual to a constant
and therefore this action can also be obtained just by dualizing $H$.
A similar analysis starting in ten dimensions can be found in
\cite{Sachs}.
\section{$AdS_3$ geometry and BTZ black holes}

For later convenience we will review in this section some
parametrizations of $AdS_3$ and discuss the discrete identifications,
which yield the BTZ black hole.

\subsection{Parametrizations of $AdS_3$}

\renewcommand{\arraystretch}{1.2}

A three-dimensional
anti-de Sitter space-time is defined as a hyperboloid in a 4-d space
with the signature $(-++-)$, i.e.
\be430
-(X^0)^2 + (X^1)^2 + (X^2)^2 - (X^3)^2 = -l^2
\ee
On the other hand a de-Sitter space is related to $\Lambda^2 = -l^2 > 0$,
i.e.\ formally to an imaginary $l$. At the same time this space defines
the $SL(2, {\rm \bf R})$ group space, i.e.\ any $g\in SL(2, {\rm \bf R})$
can be given by
\renewcommand{\arraystretch}{1.2}
\be440
g = { 1 \over l}
        \left(\ba{cc} X^0 + X^1 & X^2 - X^3 \\ X^2 + X^3 & X^0 - X^1
        \ea \right) =
{ 1 \over l}
        \left(\ba{cc} X_+ & U \\ V & X_-
        \ea \right) \ .
\ee
The $SL(2, \bf R)$ algebra reads
\be450
 [ T_a , T_b ] = \epsilon_{ab}^{ \ \  c} T_c, \hspace{0,5cm}
 {\rm Tr} (T_a T_b) = \frac{1}{2} \eta_{ab}
\ee
with the generators $T_a$ and $\eta= {\rm diag}(-1,1,1)$,
$\epsilon^{012}=1$.  A representation is given by
\renewcommand{\arraystretch}{1.2}
\be460
T_0 = \frac{1}{2} \
\left (
   \begin{array}{cc}
   0     & -1   \\
   1     & 0  \\
\end{array}
\right ), \ \ \ \
T_1 = \frac{1}{2} \
\left (
   \begin{array}{cc}
   0     & 1   \\
   1     & 0  \\
\end{array}
\right ), \ \ \ \
T_2 = \frac{1}{2} \
\left (
   \begin{array}{cc}
   1    & 0   \\
   0    & -1  \\
\end{array}
\right ).
\ee
Using these generators we can write the group element $g$
and their inverse as
\be470
\ba{rcl}
g &=&{1 \over l} \left( X^0 \, {\bf 1} \ + \ 2 X^3 \, T_0 \ +
        \ 2 X^2 \, T_1 \ + \  2 X^1 \, T_2 \right) \\
g^{-1} &=&{1 \over l} \left( X^0 \, {\bf 1} \ - \ 2 X^3 \, T_0 \ -
        \ 2 X^2 \, T_1 \ - \  2 X^1 \, T_2 \right) \ .
\ea
\ee
Next, we want to embed the string solution that we discussed before.
One way to do this is to put the horizon given by $r=0$ at the lightcone
direction $X_+ = 0$, which is related to the coordinate identifications
\be480
U = {ur \over l} \qquad , \qquad V = {vr \over l}
\qquad {\rm and} \qquad X_+ = r \ .
\ee
(so the string worldsheet $(u,v)$ is in the $(U,V)$ plane).
In 4 dimensions the metric is flat, i.e.
\be490
ds^2 = -(dX^0)^2 + (dX^1)^2 + (dX^2)^2 - (dX^3)^2 = - dX_+ dX_- + dU dV\ .
\ee
Using the constraint (\ref{430}) we can substitute the $X_-$ coordinate
and get the 3-d metric
\be500
ds^2 = \Big({r \over l}\Big)^2 \, dudv + \Big({ l \over r}\Big)^2
        \, dr^2
\ee
which coincides with the asymptotic $AdS_3$ vacuum of the BTZ black hole,
see (\ref{200}). In a more general setup, we can replace $u \rightarrow
\theta_L l $, $v \rightarrow \theta_R l$ and $r \rightarrow
e^{\lambda} l$, i.e.
\be510
V = \theta_L \, e^{\lambda}\, l  \qquad , \qquad
U = \theta_R \, e^{\lambda}\, l
\qquad {\rm and} \qquad X_- = e^{\lambda} \, l
\ee
and the $SL(2, {\rm \bf R})$ group element becomes
\be520
g =  \left(\ba{cc} 1 & 0 \\ \theta_L & 1 \ea \right)
        \left(\ba{cc} e^{\lambda} & 0 \\ 0 & e^{- \lambda} \ea \right)
        \left(\ba{cc} 1 & \theta_R \\ 0 & 1 \ea \right) =
        \left(\ba{cc} e^{\lambda} & \theta_R \, e^{\lambda} \\
        \theta_L \, e^{\lambda} & e^{-\lambda} + \theta_L \theta_R \,
        e^{\lambda} \ea \right)
\ee
There is a second parametrization that will
become important later, where one introduces polar coordinates
for the $(X^0, X^2)$ and $(X^1, X^3)$ planes
\be531
\ba{lll}
X^0 = R \cosh\theta_1 & , &
        X^1 =  \sqrt{R^2 -l^2} \cosh\theta_2
\\
X^2 = R \sinh\theta_1 & , & X^3 = \sqrt{R^2 -l^2} \sinh\theta_2 \ .
\ea
\ee
To be more precise, this parametrization is only valid in the region
$$
-(X^0)^2 + (X^2)^2 < 0
$$
The surface
$$
-(X^0)^2 + (X^2)^2 = 0
$$
is a null surface which separates the space into two regions.
In the case that
$$
-(X^0)^2 + (X^2)^2 = -R^2> 0
$$
we have to replace $R$ in the parametrization by $\sqrt{-R^2}$.
Similarly, in (\ref{531})
$$
(X^1)^2 - (X^3)^2 = R^2 - l^2
$$
For the regions where $R^2 - l^2 <0$, we have to replace
$\sqrt{R^2-l^2} $ by $\sqrt{l^2-R^2}$.
\\
Another way would be to take polar
coordinates for the two Euclidean planes $(X^0 , X^3)$ and $(X^1, X^2)$.
The space is then parametrized in the following way:
\be532
\ba{lll}
X^0 = l \cosh \lambda \sin \theta_1 & , &
        X^1 =  l \sinh \lambda \sin \theta_2
\\
X^2 = l \sinh\lambda \cos \theta_2 & , & X^3 = l \cosh\lambda\cos \theta_1.
\ea
\ee
In doing so the complete $AdS_3$ space is covered.  In this parametrization the closed
timelike curves are visible. To avoid them, one usually considers the covering
space of $AdS_3$. As before we can calculate the resulting 3-d metric and
find
\be550
\ba{rcl}
ds^2 &=& l^2 \left( \sinh^2 \lambda  \, d\theta_2^2 - \cosh^2\lambda
         \, d\theta_1^2 + d\lambda^2  \right) \\
  &=&  + \frac{R^2}{l^2} \  dy^2
- \left ( \frac{R^2}{l^2} + 1 \right ) \ dt^2  +
  \left ( \frac{R^2}{l^2} +
   1 \right )^{-1} \ dr^2     ,
\ea
\ee
where we have identified $R=l\sinh \lambda$, $ l \theta_2 = y$ and
$l\theta_1=t$.
This metric coincides with (\ref{500}) in the limit $R
\rightarrow \infty$. Thus, the string world sheet is now along the
polar angles $\theta_{1/2}$ and  in contrast to the case discussed
above it has been rotated (before the worldsheet was only in
the $X^{2/3}$ plane, see eq.\ (\ref{480})).
\\
Taking (\ref{531}), the group element $g$ in (\ref{440}) becomes
\be540
g =     e^{\theta_L T_1} \; e^{ \lambda T_2} \; e^{ \theta_R T_1} \ = \
        \left(\ba{cc} \cosh{\theta_L \over 2} & \sinh{\theta_L\over 2} \\
        \sinh{\theta_L \over 2} & \cosh{\theta_L \over 2} \ea \right)
        \left(\ba{cc} e^{\lambda/2} & 0 \\ 0 & e^{- \lambda/2} \ea \right)
        \left(\ba{cc} \cosh{\theta_R \over 2} & \sinh{\theta_R \over 2} \\
        \sinh{\theta_R \over 2} & \cosh{\theta_R \over 2} \ea \right)
\ee
where $\theta_{R/L} = (\theta_1 \pm \theta_2)$.
It is also useful to calculate the $SL(2, \bf R)$ currents,
using (\ref{470}) we find
\be552
\ba{r}
g^{-1} dg = {2 \over l^2} \, [X^0 dX^3 - X^3 dX^0 + X^1 dX^2
        - X^2 dX^1] \, T_0 \\
     + {2 \over l^2} \, [X^0 dX^2 - X^2 dX^0 + X^1 dX^3 - X^3 dX^1] \, T_1 \\
     + {2 \over l^2} \, [X^0 dX^1 - X^1 dX^0 + X^3 dX^2 - X^2 dX^3] \, T_2
\ea
\ee
or, if we express it in terms of the lightcone coordinates
($T_{\pm} = T_1 \pm T_0$)
\be553
\ba{r}
g^{-1} dg = {1\over l^2} (X_+ dV - V dX_+) T_+ + {1 \over l^2}
        (X_-dU - U dX_-)T_-\\
        + {1\over l^2}( X_- dX_+ - X_+ dX_- \; + \; VdU - UdV ) T_2 \ .
\ea
\ee
Inserting the coordinates (\ref{531}) this becomes
\be554
\ba{rcl}
g^{-1} dg &=&
   \, [-\sinh \theta_L \; d\lambda + \sinh \lambda \,
        \cosh  \theta_L \; d\theta_R ] \, T_0  \\
  &+& \, [\cosh \lambda \; d\theta_R + d\theta_L ] \, T_1 \\
  &+&  \, [\cosh  \theta_L \; d\rho - \sinh \lambda \,
    \sinh  \theta_L \; d\theta_R ] \, T_2
\ea
\ee

\renewcommand{\arraystretch}{1.7}
\subsection{The BTZ black hole as a topological solution in $AdS_3$}

How can we construct black holes in anti de Sitter space?
We know that space-time is locally anti-de Sitter. In particular,
its curvature is constant. The black hole can differ from
$AdS$ only in its global properties. It was shown in
\cite{172} that one can obtain three dimensional black holes
from the universal cover of $AdS_3$ by
dividing out by a discrete symmetry
group. The symmetry is given by a discrete subgroup
given by a particular Killing vector $\zeta$. The Killing
vectors in $AdS_3$ space generate the isometries of $AdS$.
These are the boosts in the (0,1), (2,3), (1,3) and (0,2) planes
and rotations in the (0,3) and  (1,2) planes.
The boost generators are of the follwing form
$$
J_{ab} = x_a \partial_b + x_b \partial_a,
$$
where $ab$ = $\{01,23,13,02\}$.
Rotations are generated by the vectors
$$
L_{mn} = x_m \partial_n - x_n \partial_m,
$$
where $mn$ = $03, \, 12$.
It is known that $SO(2,2) \sim SL(2,R)_L \times SL(2,R)_R$.
In terms of the generators, one $SL(2,{\bf R}) $
is generated by
$$
J_{02} - J_{13}, \quad J_{01} + J_{23}, \quad L_{03} - L_{12}
$$
and the other one is generated by
$$
J_{02} + J_{13}, \quad J_{01} - J_{23}, \quad L_{03} + L_{12}
$$
Ref. \cite{172}
obtained the black hole solutions using Killing vectors, which
are linear combinations of the boost generators in the (0,2)
and (1,3) plane. The particular form of the linear combination
determines the locations of the horizons  $r_{+}$ and $r_-$.
We will therefore pick a parametrization of the $AdS$ space,
in which the Killing vector takes a particularly simple form.
\\
Let us turn to our concrete parametrization of $AdS$ given in the
previous section. Computing the metric in the new coordinates
$R, \theta_1, \theta_2$ as introduced in (\ref{531}) yields
\begin{equation}
ds^2 = \left( \frac{R^2}{l^2} - 1 \right) ^{-1} dR^2
+ l^2 \left( 1 - \frac{R^2}{l^2} \right) d\theta_2^2 + R^2 d\theta_1^2
\end{equation}
Note that $\theta_1 $ and $\theta_2 $ are boost parameters,
i.e. $\theta_{1,2} \in \{ - \infty, + \infty \}$.
The above metric formally looks like a black hole
metric if we identify
$$
l \theta_2 = t
$$
as a time coordinate. We would also like to interpret the parameter
$\theta_1$ as an angular coordinate. However, it has the wrong
range of parameter. We have to identify
$$
\theta_1 = \theta_1 + 2 \pi
$$
(this is also necessary to avoid the conical singularity at $R=0$).
This means that we have divided out the space by a discrete symmetry.
The Killing vector corresponding to this symmetry is given by
$$
\zeta = \frac{\partial}{\partial \theta_1}
$$
This means that we divide out by the following finite
symmetry transformation
$$
e^{2 \pi \frac{\partial}{\partial \theta_1}} P \sim P,
$$
where $P$ is a point of space-time. The effect of the operation is
that $\theta_1$ has periodicity $2\pi$. We can change the periodicity by
dividing out by
$$
e^{n \pi \frac{\partial}{\partial \theta_1}} P \sim P.
$$
If we introduce the coordinates $\theta_{L/R}$, we see
that the Killing vectors $J_{01} \pm J_{23}$ are given by
$\frac{\partial}{\partial \theta_L} $ and
$\frac{\partial}{\partial \theta_R}$. For the other generators
in $SL(2,R)_L \times SL(2,R)_R $ we obain more complicated
expressions in terms of these coordinates.
\\
This is not the only way to make one of the coordinates
periodic. In fact, we can periodically identify a linear
combination of $\theta_1$ and $\theta_2$. In the
resulting  black hole solution this corresponds to adding
angular momentum. Because the metric for our brane
configuration is of the form
(\ref{170}), we are particularly interested in that case.
Let us perform the coordinate transformation
\begin{equation}
R^2 = l^2 \frac{r^2 - r_-^2}{r_+^2 - r_-^2},
\qquad
\left( \begin{array}{c} \theta_1 \\ \theta_2 \end{array} \right) =
\left( \begin{array}{cc} r_+/l^2  & -r_-/l^2 \\
                         -r_-/l^2 & r_+/l^2 \end{array} \right)
\left( \begin{array}{c} y \\ t \end{array} \right)
\end{equation}
The determinant of the matrix vanishes for $r_+= r_-$ and,
therefore, we should restrict to the case $r_+ \neq r_-$
(the extremal case is not included in this discussion).
The new metric is given by (\ref{170}). Again, both $t$ and $y$
take values on the whole real axis. To turn the metric into
a black hole metric, we have to periodically identify $y$.
That means, we choose the Killing vector $\partial_y$
and divide out by
$$
e^{n \pi \frac{\partial}{\partial y}}
$$
Note that $y$ is a linear combination of the original
$\theta$-coordinates, which means that we have rotated
the compact direction.
In terms of the theta coordinates, the Killing vector
reads
$$
\frac{\partial}{\partial y} =
\frac{r_-}{r_+^2 - r_-^2} \frac{\partial}{\partial \theta_2} +
\frac{r_+}{r_+^2 - r_-^2} \frac{\partial}{\partial \theta_1}
$$
Once again, we obtain $r_+ \neq r_-$.


\section{Chern-Simons theory and WZW models}

Now we will discuss the different CFT's in detail. In order
to do so we re-write, first of all,
the BTZ black hole as a Chern-Simons theory. The
spatial part of the geometry is given by an annulus and on both
boundaries live different CFT's. In two subsections we analyse the two
boundaries separately.
\subsection{The BTZ black hole as a Chern-Simons model}
It is known that Einstein-anti-de Sitter gravity in $2+1$ dimensions,
as given in eq.\ (\ref{390}), is equivalent to Chern-Simons theory
\cite{Achucarro} (for a discussion of an additional matter part
see \cite{551}). Choosing conventions where the
three-dimensional gravitational coupling is related to the level $k$ by
\be572
 k = \frac{2 \pi l }{\kappa^2_3} = {p^2 p^3 \over \alpha'}
\ee
and decomposing the diffeomorphism group $SO(2,2) \simeq SL(2,{\bf
R})_L \times SL(2, {\bf R})_R$ the 3-dimensional action can be written as
\be560
S = S_{CS} [A] \ - \  S_{CS} [\bar A]
\ee
with
\be570
S_{CS} [A] = \frac{k}{4 \pi} \int_{M_3} d^3 x
        {\rm Tr} \ (AdA + \frac{2}{3} A^3 ) \ .
\ee
The gauge field one-forms are
\be580
A = (\omega^a + \frac{1}{l} e^a) \ T_a \ \in SL(2,{\bf R})_R \ , \hspace{1cm}
\bar A = ( \omega^a - \frac{1}{l} e^a) \ \bar T_a \ \in SL(2,{\bf R})_L.
\ee
where $\omega^a \equiv \frac{1}{2} \epsilon^{abc} \omega_{bc}$ are given by
the spin-connections $\omega_{bc}$ and $e^a$ are the dreibeine. Under gauge
transformations
\be590
A \rightarrow g^{-1} (A + d) g
\ee
the Chern-Simons action transforms as
\be600
S_{SC}[A] \rightarrow S_{SC}[A] - {k \over 12} \int_M (g^{-1}dg)^3 -
        {k \over 8\pi} \int_{\p M} \Big[ (g^{-1}dg)_v (g^{-1}Ag)_u -
        (g^{-1}dg)_u (g^{-1}Ag)_v \Big]
\ee
where the integral over $\partial M$ comprises all boundaries. The model
is therefore gauge invariant if (i) there are no boundaries or (ii) if
the gauge field are trivial on the boundaries and the topological charge
coming from the $(g^{-1}dg)^3$ term is integer-valued. However, anti-de
Sitter spaces have boundaries and the fields do not vanish there.
Furthermore, $AdS$-spaces are globally not hyperbolic. Thus, to
obtain a reliable theory, one has to impose boundary conditions \cite{571}
(see also \cite{590}).
As a consequence gauge degrees of freedom do not
decouple and become dynamical at the boundaries. These are the degrees of
freedom of the conformal field theories living at the boundaries.
\\
In the following we will discuss this procedure for the BTZ black hole.
The geometry of the manifold is $M_3 = {\bf R} \times \Sigma$, where
${\bf R}$ corresponds to the time of the covering space of $AdS_3$ and
$\Sigma$ represents an ``annulus'' $r_+ \leq r < \infty$.
\\
For the metric (\ref{170}) the dreibeine are ($ds^2 = - e^0 e^0
+ e^1 e^1 + e^2 e^2$)
\be610
e^0 = e^{-V} dt \quad , \quad
e^y = \Big({r \over l}\Big)\, \Big(dy - {r_- r_+ \over r^2} \, dt \Big)
\quad , \quad
e^r = e^{V} dr
\ee
and using the relation
$$
de^a + \omega^a_{\quad b} \wedge e^b = 0
$$
one obtains for the spin-connections
\be620
\ba{l}
\omega^{0r} = e^{V} {r \over l^2} (1 + {r_- r_+ \over r^2}) \; e^0
        -{r_- r_+ \over r^2 l} \; e^y  \\
\omega^{yr} = e^{-V} {1 \over r} \; e^{y} + {r_- r_+ \over  r^2 l} \; e^0 \\
\omega^{0y} = - {r_- r_+ \over r^2 l} \; e^{r} \ .
\ea
\ee
It follows that the gauge connections $A= A^a T_a$ and $\bar A =
\bar A^a \bar T_a$ are given by
\be630
\begin{array}{ll}
  A^0 = e^{-V} \ {dv \over l} \ , \hspace{1cm}     &
  \bar A^0 = e^{-V} \ {du \over l} \ ,      \\
   A^1 = {r \over l} (1 - {r_- r_+ \over r^2}) \ {dv \over l} \ ,
        \hspace{1cm} &
  \bar A^1 = - {r \over l} (1 + { r_- r_+ \over r^2}) \ {du \over l} \ ,  \\
   A^2 = e^{V} (1 + {r_- r_+ \over r^2})  \ \frac{dr}{l}\ , \hspace{1cm} &
   \bar A^2 = - e^{V} (1 - {r_- r_+ \over r^2}) \frac{dr}{l}    \\
\end{array}
\ee
or, equivalently,
\be640
\begin{array}{l}
A = \Big( e^{-V} \, T_0  + {r \over l} (1 - {r_- r_+ \over r^2}) \, T_1 \Big)
        {dv \over l} + e^{V} (1 + {r_- r_+ \over r^2})\, T_2  \
        \frac{dr}{l} \ , \\
\bar A = \Big( e^{-V} \, T_0  - {r \over l} (1 + {r_- r_+ \over r^2}) \,
        T_1 \Big) {du \over l}  - e^{V} (1 - {r_- r_+ \over r^2}) \,
        T_2 {dr \over l} \ .
\end{array}
\ee
These fields are pure gauges ($F=\bar F =0$).
Performing particular coordinate transformations $A$ becomes
\be642
A = g^{-1} dg =
\Big({r_+-r_-\over l} \sinh\lambda \; T_0 + {r_+ - r_- \over l}
   \cosh\lambda \; T_1\Big)\, {dv \over l} \; + \; T_2 \, {d \lambda
   \over l}
\ee
where $\sinh \lambda = \, {l \over r_+ - r_-} \, e^{-V}$ and $g$ is
given by (\ref{540}) with $\theta_L = 0$ and $\theta_R = v$ (compare
also with (\ref{554})). Analogous one obtains for $\bar A$
\be644
\bar A = \bar g^{-1} d\bar g =
   \Big(- {r_+ + r_- \over l} \, \sinh\lambda \; T_0 - {r_+ + r_-
    \over l}\, \cosh\lambda \; T_1 \Big)\, {du \over l} \; - \; T_2 \,
    {d \lambda \over l}
\ee
where $\sinh \lambda = \,- {l\over r_+ + r_-} \, e^{-V}$ and
$\bar g$ is again given by (\ref{540}), but now with $\theta_L = 0$ and
$\theta_R = u$.
\\
Obviously any rescaling of the form
$g \rightarrow g_0 g$ with a constant
group element $g_0$ gives an equivalent parametrization and, thus,
the group elements $g$ and $\bar g$ are not uniquely fixed.
\\
At the boundaries the gauge fields take the value
\be650
\ba{rl}
{\rm for}\ r \rightarrow \infty:\quad &
  A \ = \ {r\over l} (T_1 + T_0)
  \; {dv \over l} + T_2 \; {dr \over r} \ = \ {r\over l} \, T_+
  \, {dv \over l} + T_2 \; {dr \over r}
  \\
  & \bar A \ =\ - {r\over l} (T_1 - T_0) \; {du \over l} - T_2 \; {dr \over r}
  \ = \ - {r\over l} \, T_- \, {du \over l} - T_2 \; {dr \over r} \ .
\ea
\ee
with $T_{\pm} = T_1 \pm T_0$.  In terms of $\lambda$,
the horizon boundary $r\rightarrow r_+$ is mapped to $\lambda
\rightarrow 0$ and the gauge fields become
\be670
\ba{rl}
{\rm for}\ \lambda \rightarrow 0 \quad ({\rm or}\ r \rightarrow r_+)  :
 \quad &
  A \ = \ {1 \over l} (r_+ - r_-) \; T_1  \; {dv \over l} +
  \; T_2 \; d\lambda \\
  & \bar A \ =\ - {1\over l}(r_+ + r_-) \; T_1 \; {du \over l} -
  \; T_2 \; d\lambda \ .
\ea
\ee
But these gauge fields do not follow from the variational principle
for the Chern-Simons action (\ref{570}), which yields
\be680
\delta S_{CS}[A] = {k \over 2 \pi} \int_M \delta A \wedge F - {k\over 8 \pi}
        \int_{\partial M} \left[A_v \delta A_u - A_u \delta A_v \right]\ .
\ee
The vanishing of the bulk variations means that the field strength
has to be zero (i.e.\ pure gauge), which is in fact the case for the
BTZ solution. This statement holds also at the quantum
level, where one allows for arbitrary gauge fields, i.e.\ not only
classical solutions. Namely, as consequence of our geometry $M_3 = {\bf
R} \times \Sigma$ (where ${\bf R}$ corresponds to the time), the time
component of the gauge field $A_0$ appears as a Lagrange multiplier in
the action and integrating out this Lagrange multiplier from the
quantum effective action
yields the constraint $F_{yr} =0$ \cite{440}. Hence, at the
quantum level the connections on $\Sigma $ are also flat.
\\
On the other hand, treating boundary variations in the same (independent)
way as bulk variations yields $A_u = A_v = 0$ at $\p M$, i.e. gauge
transformations have to vanish at the boundaries. However, this is not
the case for our solution (\ref{640}), which has
non-trivial boundary values presented in (\ref{650}) and (\ref{670}). A
simple way to obtain non-trivial gauge fields at the boundaries from the
action principle is to add further terms.
\\
In the following we will discuss both boundaries separately.

\subsection{The CFT at the asymptotic boundary}

In order to obtain the correct CFT at the boundary, one has to
take two points into account:

(i) To comply with the action principle we have to add
additional boundary terms to the action.

(ii) The CFT is given by the gauge degrees of freedom that become
dynamical on the boundary, i.e. the CFT is related to the broken gauge
symmetries. However, the BTZ solution has still an invariant subgroup and
therefore the CFT does not correspond to the complete $SL(2,\bf R )$
group, but to an $SL(2, \bf R )$-coset,
where the invariant subgroup is modded out.

We will start with the first point. As suggested by our classical solution we
will consider the following boundary conditions at infinity
\be690
A_u = \bar A_v = 0 \ .
\ee
To obtain the correct boundary conditions in agreement with the variational
principle one has to add additional boundary terms to the
quantum effective action. Considering
\be700
\ba{rcl}
S[A,\bar A ] &=& S_{CS}[A]  + B_{\infty}[A] - S_{CS}[\bar A] -
  B_{\infty}[\bar A] \\
  &=& {k \over 4 \pi} \int_M (AdA + {2 \over 3} A^3)
  - {k \over 4 \pi} \int_M (\bar A d\bar A + {2 \over 3} \bar A^3) +
    {k \over 8 \pi} \int_{\p M_{\infty}} (A_v A_u + \bar A_u \bar A_v)
\ea
\ee
yields the variation
\be710
\delta S = { k \over 2 \pi} \int_{M} \delta A \wedge F -
     { k \over 2 \pi} \int_{M} \delta \bar A \wedge \bar F
     + {k \over 4\pi} \int_{\p M_{\infty}} (A_u \delta A_v + \bar A_v
     \delta \bar A_u)
\ee
and, therefore, since $\delta A$ are arbitrary, one obtains the
following equations of motion and boundary conditions
\be720
\ba{rl}
F = \bar F = 0 & \qquad {\rm in } \quad M \\
A_u = \bar A_v= 0 & \qquad {\rm on} \quad \p M_{\infty} \quad (r=\infty)
\ea
\ee
which is in agreement with (\ref{690}).
\\
As argued below (eq.\ (\ref{680})), the field strength has to vanish at the
classical and at the quantum level and therefore we can write
\be730
A = g^{-1} dg  \qquad {\rm and} \qquad \bar A = \bar g^{-1} d\bar g \ .
\ee
Inserting these fields into the action (\ref{700}) one obtains two chiral
WZW models (due to the boundary terms). Combining both chiral models to
one non-chiral WZW model yields \cite{573}
\be740
\ba{rcl}
S[A, \bar A] &=&  S_{cWZW_v}[ g^{-1}] + S_{cWZW_u}[\bar g] =
S_{WZW}[\, \hat g^{-1} ] \\
&=& {k \over 4 \pi} \int_{\p M} tr (\hat g^{-1} d \hat g)
        (\hat g^{-1} d \hat g) - {k \over 6\pi} \int_{M} tr
        (\hat g^{-1} d \hat g)(\hat g^{-1} d \hat g)(\hat g^{-1} d \hat g) \ .
\ea
\ee
where $\hat g = g \bar g^{-1}$ can be parametrized by
\renewcommand{\arraystretch}{1.2}
\be750
\hat g =  \left(\ba{cc} 1 & 0 \\ \theta_L & 1 \ea \right)
        \left(\ba{cc} e^{\lambda} & 0 \\ 0 & e^{- \lambda} \ea \right)
        \left(\ba{cc} 1 & \theta_R \\ 0 & 1 \ea \right) \ .
\ee
In order to make this result more transparent it is useful to consider
the classical solution (\ref{650}), for which the gauge group elements read
\be760
g =  \left(\ba{cc} 1 & 0 \\ \theta_L  & 1 \ea \right)
        \left(\ba{cc} e^{\lambda/2} & 0 \\ 0 & e^{- \lambda/2} \ea \right)
\quad , \quad
\bar g =  \left(\ba{cc} 1 & \theta_R \\ 0 & 1 \ea \right)
        \left(\ba{cc} e^{-\lambda/2} & 0 \\ 0 & e^{ \lambda/2} \ea \right)
\ee
with $e^{\lambda} = {r \over l}$,  $\theta_L = {v \over l}$ and
$\theta_R = {u \over l}$. Combining both elements yields (\ref{750}).
\\
As mentioned at the beginning of this section, the boundary CFT is not
a complete $SL(2, \bf R)$ model but a coset model. In order to find
the correct coset one has to determine the invariant subgroup.
Examination of the group elements of the BTZ model (\ref{760})
yields that gauge transformations of the type
\be770
g \rightarrow \left(\ba{cc} 1 & 0 \\ \alpha_L  & 1 \ea \right) \; g
\qquad {\rm and} \qquad
\bar g \rightarrow \left(\ba{cc} 1 & \alpha_R \\ 0 & 1 \ea \right) \; \bar g
\ee
\renewcommand{\arraystretch}{1.7}
can be absorbed into a redefinition of $u$ and $v$. Note that
both gauge connections $A_{\mu}$ and $\bar A_{\mu}$
do not depend on these two coordinates, which
correspond to Killing vectors.
For the CFT this symmetry means, that as group space we have to consider
a coset model and, therefore, the WZW models must be gauged.
For the model at hand one has to gauge the group directions
generated by $T_{\pm}$ which is given by \cite{162}
\be780
S_{WZW} \rightarrow S_{WZW} + {k \over 2 \pi} \int_{\p M_{\infty}}
\left[ a_v (e^{2 \lambda} \p_u \theta_L - \sqrt{\mu}) + a_u(
e^{2 \lambda} \p_v \theta_R - \sqrt{\mu}) + a_u a_v e^{2 \lambda}
\right] \ .
\ee
It is invariant under $\theta_{L/R} \rightarrow \theta_{L/R} +
\alpha$, $a \rightarrow a + d \alpha$.  Integrating out the gauge
fields $a_{u/v}$ yields
\be790
S = {k-2 \over 4 \pi} \int_{\p M_{\infty}} \left[ \partial_u \lambda
\p_v \lambda  + Q R^{(2)} \lambda + \mu \, e^{-2 \lambda} \right]
\ee
which is the Liouville model. This gauged model is equivalent
to keeping fix the currents $ J_{\pm}
= \sqrt{\mu}$ (see \cite{422}) and thus the value of the Liouville mass
parameter $\mu$ can be matched with the classical boundary values
appearing in (\ref{650}), i.e.\
one may take\footnote{Note that like $\mu$ also $l$ is an undetermined
quantity; due to the scaling symmetry of the asymptotic vacuum
$r\rightarrow \rho r$ and $l\rightarrow \rho l$. Hence, in a quantum
theory these are ``bad'' expansion parameters.} $\sqrt{\mu} = 1/l$.
The shift $k \rightarrow k-2$ is a renormalization effect and in
supersymmetric models one may undo this shift. Finally
the background charge $Q$ comes from performing the Gaussian
integral. Equivalently, the appearance of this term
is required by conformal invariance
and even for vanishing 2-d curvature ($R^{(2)} = 0$) one has to take
into account the background charge $Q$. To make this connection more
clear let us mention, that $\lambda$ corresponds to the radial
coordinate of the target space and the Liouville model can be seen as
a $\sigma$-model description of two scalar fields, the dilaton
$\phi(\lambda)$ and a tachyon $T(\lambda)$
\be792
S = {k-2 \over 4 \pi} \int_{\p M_{\infty}} \left[ \partial_u \lambda
\p_v \lambda  + Q R^{(2)} \phi(\lambda) + T(\lambda) \right] \ .
\ee
This model is conformal invariant at the quantum level, if the
corresponding $\bar \beta$-functions vanish \cite{575},
which are interpreted as equations of motion for these scalar fields
\be794
\partial^2 \phi = 0 \qquad , \qquad
- {1\over 2(k-2)} \partial^2 T -2 T
+ \partial \phi \, \partial T = 0
\ee
Taking the fields from (\ref{790}), the first equation is solved
trivially by the linear dilaton $\phi = Q \lambda$. In the second
equation we insert $T \sim e^{-2\lambda}$ and find for $Q$
\be795
Q= {1-k \over k-2} \ .
\ee
By redefining $T\rightarrow e^{(k-2) \phi}\, T$ the second equation
becomes the Klein-Gordon equation
\be796
\Big(\p^2   - (k-3)^2 \Big) T = 0 \
\ee
which is massless for $k=3$. As already mentioned in the supersymmetric
case we have to undo the shift in $k$, i.e.\ we have
to replace $k \rightarrow k+2$ and the massless point corresponds to
$k=1$. On the other hand, $k$ was introduced as the radius of the
$S_3$ space measured in $\alpha'$, see (\ref{150}). When expressed in
terms of the number of 5-branes ($m$) and KK-monopoles ($n$), $k = mn$
and the massless case correspond to a single 5-brane and KK-monopole.
\\
Finally, we have to determine the central charge. An easy way to do
this, is to calculate the dilaton-$\bar \beta$ function, which
gives  as consequence of Zamolodchikov's c-theorem
the central charge ($\bar\beta^{\phi} = {c \over 6}$ at the conformal
fixpoint, see \cite{575}). We find
\be800
c_L = 1 + 6\, (k-2)\, Q^2 = {3k \over k-2} -2 + 6k  \ .
\ee
where ${3k \over k-2}$ is the $SL(2,\bf R)$ central charge; the
``-2'' is due to the fact, that both $\theta$ coordinates have been
gauged away and the last $6k$ contribution corresponds to the
improvement term in the energy momentum tensor.
In the classical limit ($k \rightarrow \infty$) only the last term
contributes and yields $6k$. Moreover,  the central charge
is invariant under the transformation
\be802
k-2 \rightarrow {1 \over k-2}
\ee
and $k=3$ is just the self-dual point.  This point coincides with the
massless case and we find for the central charge $c_L = 25$, which
corresponds to the famous $c_m =1$ barrier in non-critical string
theory.  So, at this point we have to expect a phase transition.
\\
Again taking the shift $k \rightarrow k+2$ for the supersymmetric case
the symmetry transformation becomes $k \rightarrow {1 \over k}$.
Since $k= {l^2 \over 4 \alpha'}$ it can also be written as
\be804
l \ \rightarrow \ {4 \alpha' \over l} \qquad {\rm or} \qquad
\sqrt{\alpha'} \ \rightarrow \ {l^2/4 \over \sqrt{\alpha'}} \ .
\ee
So it appears as a kind of $T$-duality for the cosmological
constant $l$ or, keeping fix the cosmological constant, it is some
kind of strong-week duality (S-duality) in the $\alpha'$ expansion.
But one has to keep in mind, that although it is a symmetry of
$AdS_3$ gravity it is not a symmetry of our string inspired model,
where $k$ has to be integer-valued.
\\
In non-critical string theory this symmetry is subtle,
because the Liouville vertex operator appears as conformal factor of
the 2-d worldsheet metric and the self-dual point corresponds to a
puncture of the worldsheet. However, as discussed in \cite{580} beyond
this point ``small area divergencies'' appear related to non-normalizable
states, which spoil the worldsheet interpretation.
It is unclear to us to which extend these
objections hold in our setup, see also \cite{581}.

There are interesting lines for continuations, e.g.\
it would be interesting to add to the tachyon field additional conformal
matter or to use the procedure described in \cite{200}
to integrate out the Liouville field $\lambda$ and to obtain the
partition function and to calculate amplitudes.
\subsection{The CFT at the horizon boundary}

We proceed analogous to the case discussed above. Again
we consider the boundary condition (\ref{690}), but this time we have
to take into account a different isometry group, which can be determined
by analysing the BTZ model as given in (\ref{642}) and
(\ref{644}). The corresponding gauge group elements are given by
\be820
g =   e^{ \lambda T_2} \; e^{ ({r_+ -r_- \over l^2}) v \, T_1}
\qquad ,\qquad
\bar g =  e^{ - \lambda T_2} \; e^{({r_+ + r_- \over l^2}) u \, T_1} \ .
\ee
The isometries of the BTZ black hole correspond again to
reparametrizations of $u$ and $v$ corresponding to the
gauge transformation
\be830
g \rightarrow g \; e^{\alpha \, T_1} \qquad {\rm and} \qquad
        \bar g \rightarrow \bar g \; e^{\alpha \, T_1} \ .
\ee
The crucial difference to the CFT at the asymptotic boundary is, that
the group direction has changed, which corresponds now to deformations
in the $T_1$ direction. Note that using the identity
\renewcommand{\arraystretch}{1.1}
\be832
T_1 = g_0^{-1} \, T_2 \, g_0 \qquad , \qquad g_0 = {1 \over \sqrt{2}}
\left(\ba{cc} 1 & 1 \\ -1 & 1 \ea \right)
\ee
\renewcommand{\arraystretch}{1.7}
one can replace everywhere $T_1$ by $T_2$, i.e.\ both directions are
equivalent. After combining both chiral WZW models as in (\ref{740}),
we have to mod out this direction. The corresponding gauged WZW is
given by
\cite{160,162}
\be840
\ba{r}
S_{WZW} \rightarrow S_{WZW} + {k \over 2 \pi} \int_{\p M_{\infty}}
        \left[ a_v (\p_u \theta_L  + \cosh\lambda \p_u \theta_R )  + \right.\\
        \left. \qquad +
        a_u( \p_v \theta_R + \cosh\lambda \p_u \theta_R  ) -
        a_u a_v (\cosh\lambda +1)  \right] \ .
\ea
\ee
This model has been extensively studied as a model describing 2-d
black holes and it can also be written as a
$\sigma$-model
\be850
S \sim \int d^2 \xi \left[ \partial_{\alpha} X^{\mu}
        \partial^{\alpha} X^{\nu} G_{\mu\nu} + \alpha'
        R^{(2)} \phi(X) \right]
\ee
but now with a 2-d target space $X^{\mu} = \{\theta_2, \lambda\}$. In
order to obtain the metric  $G_{\mu\nu}$ one fixes the
gauge and integrates out the
gauge fields $a$ and $\bar a$ in (\ref{840}). It follows that one
$\theta$ angle drops out. But using this approach one obtains only the
lowest order metric. An alternative approach,
discussed in \cite{162}, is to consider the $L_0$ operator as a
target space Laplacian. As consequence the mass shell condition of the
tachyon vertex operator becomes an analogous Klein-Gordon equation as
given in (\ref{796}). The corresponding exact background metric and
dilaton \cite{162} read
\be860
\ba{l}
ds^2 =  2 (k-2) \left[d\lambda^2 -  B^{-2}(r) \, d\theta^2_2\right] \ , \\
        e^{-2\phi} = B(r)\, \cosh\lambda \, \sinh\lambda   \qquad , \qquad
        B^2(r) = (\coth^2 \lambda -     {2\over k})
\ea
\ee
However, the BTZ black hole solution is dilaton-free and also the metric is
asymptotically not flat. Where is the 2-d black hole then?
Following the procedure discussed in \cite{Welch} we
T-dualize the BTZ black hole (\ref{170}) over the coordinate $y$.
Keeping in mind that we have a non-zero antisymmetric tensor
$B_{0y} ={r^2 \over l^2}$ (see (\ref{370}) and remembering
that the $\epsilon$
tensor contains $\sqrt{g} = {r\over l}$) and after diagonalizing the T-dual
metric by
\[t \rightarrow {l\over \sqrt{r_+^2
-r_-^2}}(y -t) \quad , \quad y\rightarrow {1 \over l \sqrt{r_+^2 - r_-^2}}
(r_-^2 \, y - r_+^2 \, t)\]
one finds the black string solution \cite{540} (see also \cite{541})
\be870
ds^2 = - (1 - {r_+^2 \over r^2}) \, dt^2 + (1 - {r_-^2 \over r^2})\,
        dy^2 + e^{2V} dr^2 \ .
\ee
Moreover, gauged WZW-models correspond to compactifications of one direction.
So, after compactifying $y$ and transforming
\be872
e^{2V} dr^2 = l^2 d\lambda^2 \qquad {\rm with} \qquad
r^2 = r_+^2 \cosh^2 \lambda - r_-^2 \sinh^2 \lambda
\ee
one obtains
\be880
\ba{l}
ds^2 = l^2 d\lambda^2 - \tilde B^{-2}(\lambda) \, dt^2 \ ,\\
e^{-2\phi} = {\sqrt{2}(r_+^2 - r_-^2) \over l^2} \, \tilde B(\lambda)\,
   \cosh\lambda\,  \sinh\lambda  \qquad , \qquad \tilde B^2(\lambda)=
        {r_+^2 \coth^2 \lambda - r_-^2 \over r_+^2 - r_-^2} \ ,
\ea
\ee
which (up to constant rescalings) coincides exactly with the metric
(\ref{860}) from the conformal field theory. Note, that in
this 2-d model the dilaton corresponds to the (dual) compactification
radius of the string direction ($\sim g_{yy}$).
\\
It follows from the 2-d black hole solution that this result is valid
only in the non-extreme case. For the extreme case one has to make
different coordinate transformations and one does not obtain a 2-d black
hole. Instead, one finds travelling waves along an extremal string
\cite{550}. But also this model is an exact CFT \cite{560}.
\\
It is interesting to note, that already the classical model corresponds
to an exact CFT, not only in the extremal but also in the non-extremal
case. Therefore the geometry describes an exact background in all orders
in $\alpha'$; only the parameters (like $k$, the central charge or the
cosmological constant) have to be renormalized. This renormalization
is however obvious if one keeps in mind, that $\alpha'$ corrections
correspond e.g.\ to higher curvature corrections. The curvature
tensor and torsion (see (\ref{370})) of the 3-d model are given by
\be890
R_{\mu\rho\nu\lambda} = - {1 \over l^2}( g_{\mu\nu} g_{\rho \lambda}
  - g_{\mu\lambda} g_{\rho \nu}) \quad , \quad H_{\mu\nu\rho}
  ={2 \over l} \epsilon_{\mu\nu\rho} \ .
\ee
Both quantities are covariantly constant and any possible corrections
to the equations of motion (e.g.\ from $R^2$) are proportional to the
lowest order equations, because e.g.\ $R^n \sim R$ or $(R^m)_{\mu\nu}
\sim R_{\mu\nu}$ for arbitrary powers $n$ and $m$ of the curvature
tensor.  Thus, the exact equations of motion (to all orders in
$\alpha'$) have to have the same structure as the lowest order
equations\footnote{The dilaton and the tachyon of the conformal field
theories are scalars coming from the compactification. The 3-d model
is given only by the metric and antisymmetric tensor. The exactness of
this model can also be understood from the fact that the space is
paralizable, i.e.\ the generalized curvature tensor vanishes
\cite{575}.}.


\section{Comparison with results from  world-volume theory }

According to recent developments, the supergravity
on $AdS_3 \times S^3$ should be dual to a two-dimensional
superconformal field theory, which is realized as the world volume theory
of a brane. Here, we are dealing with only NS charges and the
dual conformal field theory is realized on the worldvolume
of the fundamental string. The near horizon limit
on the supergravity side corresponds to the
infra-red limit of the brane theory. If we embed our configuration
in a IIB context, we know that the theory on a IIB fundamental
string is the theory of a vector multiplet. In addition to the string,
we have a NS-5-brane in our setup. Therefore, we are in the S-dual
situation of the D5-D1 system studied in \cite{400}. The 5-branes
lead to fundamental hypermultiplets in the gauge theory. The metric
on the Coulomb branch of a $U(1)$ gauge theory (which translates
to a single string in terms of branes) with $k$
hypermultiplets is given by \cite{700}
\begin{equation}
ds^2 = |d\phi|^2 \left(\frac{1}{2e^2} + \frac{k}{2|\phi|^2} \right),
\end{equation}
where $\phi$ denotes the scalars of the vector multiplet. The first term
is the classical metric and the second term denotes a one loop contribution.
In a setup with 1-branes and 5-branes, the number of 5-branes corresponds
to the number of hypermultiplets.
Hence, the 5-brane metric is recovered from
the gauge theory. In our particular
setup we also added a magnetic monopole, so that
we are dealing with 5-branes at orbifolds. Thus, the gauge symmetry on
the 5-branes becomes $U(p_2)^{p_3}$, instead of $U(p_2)$. This is in
agreement with the form of the metric on the supergravity side,
where $p_2$ and $p_3$ always enter together as a product.
The Coulomb branch of the gauge
theory describes the motion of the string transversal to the 5-brane.
In addition, we have a Higgs branch describing the motions inside the 5-brane.
This is the relevant phase in the IR limit. The decoupling of Coulomb
and Higgs branch was interpreted in the context of Matrix theory as the
decoupling of the 5-brane theory from the bulk physics \cite{710}.
The situation of a wrapped NS-5-brane was considered in \cite{720}.
Here, it was argued that the relevant conformal field theory is a
$\sigma-$ model, whose target space is a symmetric product
of the internal space.
In our particular setup
we finally add a monopole. As a consequence the
$S^3$ is modded out by a discrete subgroup and supersymmetry is partially
broken. In a 4-dimensional context it was shown in \cite{730} that
modding out the $S^3$ on the supergravity side corresponds to ``orbifolding''
the conformal field theory on the brane. A similar procedure should be
applied in the two-dimensional case, too.

\section{Relation to the black hole entropy}
Very recently the $AdS$/CFT correspondence shed some new light
on the microscopic derivation of the macroscopic Bekenstein-Hawking
entropy \cite{520,400,410,412,510,Andy1,620,500,430,551}.
The reason is, that it is sometimes straightforward
to count states of CFTs and a lot of black hole solutions give
rise to a background metric of the form $AdS \times M$ near the
horizon, where $M$
denotes a compact space.
If one assumes that the Bekenstein-Hawking entropy should be
accounted for by microstates near the horizon, then it is obvious
that the $AdS$/CFT correspondence can play an important role in order
to find a statistical interpretation of the black hole entropy.
\\
In the BPS limit the leading part of the classical
black hole entropy coming from string theory is given by
\begin{eqnarray}
 S &=& 2 \pi \ \sqrt{\frac{1}{6} c_{tot} N},
\end{eqnarray}
where $N_L \equiv N$ denotes the number operator and
$c_{tot}$ the ``effective'' central charge of the underlying
CFTs. The oscillator number $N$ can be obtained from the level
matching condition and, for the particular setup discussed in the
previous sections, one obtains\footnote{For a detailed discussion see
\cite{412}. Note also that we take $\alpha^\prime=1$ in this section.}
\begin{eqnarray}
 N &=& 1 + q_0 q_1.
\end{eqnarray}
In order to map this heterotic result to the type II side, one has to
perform the symplectic transformation $q_1 \rightarrow p^1$.
\\
In \cite{412} it has been argued that the
effective central charge is of the general form
\begin{eqnarray}
 c_{tot} &=& 6k + \beta + \frac{\gamma}{k}
\end{eqnarray}
in the supersymmetric case. Here the first term denotes
the classical ($k \rightarrow \infty$) central charge that
comes entirely from the Liouville-model living at the
outer boundary of $AdS_3$. The constant shift,
parameterized by $\beta$, has been
calculated by Maldacena, Strominger and Witten (MSW) in
\cite{900}. In \cite{412} it has been shown that
additional subleading contributions, coming from the CFTs at the inner
boundary, i.e. the horizon of the BTZ black hole, and the
outer boundary of $AdS_3$, must be taken into account, too.
For the particular example given in \cite{412} it turned out
that the black hole entropy had a $k \leftrightarrow 1/k$ exchange
symmetry ($\gamma=6$) due to this additional subleading contributions to
the effective central charge.
\\
Since the inclusion of all CFTs should yield an exact formula for
the black hole entropy to all orders in $\alpha^\prime$, it is
challenging to test the proposal of \cite{412} for
the entropy by an independent field theoretical calculation of the
macroscopic Bekenstein-Hawking entropy including higher-order
curvature corrections.
\\
In doing so we follow the approach of \cite{910}. We choose
as an example the heterotic $S$-$T$-$U$ model \cite{920}
on $T^6$ with
$N=4$ supersymmetry.
As result we obtain:

(i) General case: The moduli obtain explicit higher-order corrections,
but the entropy contains no explicit corrections. Only the charge
$q_0$ obtains implicit higher-order corrections, i.e. $q_0$
gets ``renormalized''. The results are all consistent, but strictly
speaking the approach does not ``prove'' anything.

(ii) Special case:
At special points in moduli space one obtains a
pure (subleading) logarithmic black hole entropy \cite{930}. It follows
that higher-order curvature corrections (non-perturbative instanton
corrections) can stabilize black hole solutions.
\subsection{General formulae}
Black holes in the context
of $N=2$ supersymmetry and their corresponding entropies appeared as
solutions of the equations of motion of $N=2$ Maxwell-Einstein
supergravity action, where the bosonic part of the action contains at most
two space-time derivatives. This part of $N=2$
supergravity actions can be encoded in a holomorphic
prepotential $F^{(0)} (\hat X)$, which is a function of the scalar fields
$ \hat X$ belonging to the vector multiplets.
The $N=2$ effective action of strings and M-theory contains in
addition an infinite number of higher-derivative terms involving
higher-order products of the Riemann tensor and the vector field
stengths. A subset of these couplings in $N=2$ supergravity can be
described by a holomorphic function $F(\hat X, \hat W^2)$,
where the chiral superfield
$\hat W^2 = \hat W_{\mu\nu} \hat W^{\mu\nu}$
is the Weyl superfield \cite{940}. Its lowest component is the graviphoton
field strength (in form of an auxiliary field $T_{\mu\nu}^-$).
In the following we expand
$F(\hat X, \hat W^2)$ to first order in $\hat W^2$, i.e. we consider
perturbation theory in $\hat W^2$. In order to discuss the black hole
entropy and the stabilisation equations it is convenient to introduce
new quantities
$(X^I,W^2)= (\bar Z \hat X^I,\bar Z \hat W^2)$,
where $Z$ denotes the graviphoton charge.
Thus, we consider a general expansion of the form
\begin{eqnarray}
 F(X,W^2) &=& \sum_{g=0} \ F^{(g)}(X) \ W^{2g}
\end{eqnarray}
In the following we will not solve the equations for the full
black hole solution. Instead we will impose the stabilisation
equations.
The metric of the black hole solution is given by
\begin{eqnarray}
ds^2 &=& - e^{2U(r)} \ dt^2 \ + \ e^{-2U(r)} dx^m dx^m
\end{eqnarray}
and the metric function reads
\begin{eqnarray}
e^{-2U} &=& Z \bar Z \ = \
i \
\left (
 \bar X^I F_I - \bar F_I X^I
\right )
\end{eqnarray}
The stabilisation equations are given by \cite{991}, \cite{990}
\begin{eqnarray}
i (X^{I}- \bar X^{I}) &=&  \tilde H^{I},
\hspace{2cm}
i (F_{I}- \bar F_{I}) \ = \  H_{I}
\end{eqnarray}
with harmonic functions
\begin{eqnarray}
\tilde H^{I} &=& \tilde h^I + \frac{p^I}{r},
\hspace{2cm}
H_{I} \ = \  h_I + \frac{q_I}{r}.
\end{eqnarray}
Considering the lowest order ($g=0$),
it has been shown in \cite{991}, \cite{990} that these conditions are sufficient
for solutions of $N=2$ supergravity breaking half of $N=2$
supersymmetry,
i.e. these solutions solve the equations of motion, the Bianchi
identities and give rise to Killing spinors, such that the
corresponding background is purely bosonic.
If the charges of the harmonic functions satisfy additional
constraints, these solutions represent black holes.
The corresponding entropy of spherically symmetric black holes is given by
\begin{eqnarray}
S_{BH} &=& \lim_{r \rightarrow 0} \ \pi r^2 \ e^{-2U(r)}.
\end{eqnarray}
Another important point is to determine $W^2$ (at least on the
horizon):
As long as we are only interested in the first order correction
(linear in $W^2$) we can expand
\begin{eqnarray}
\label{def_1}
 T^-_{\mu\nu}  &=& M_I F_{\mu\nu}^{I-} - L^I G^-_{I \mu\nu}
\end{eqnarray}
with
$G^-_{I \mu\nu}= G_{I \mu\nu}^{(0)-} + G_{I \mu\nu}^{(1)-}(T^-) $, i.e.
(\ref{def_1}) is an implicit equation. However, if we
assume\footnote{Here we consider $T^-$ to remain
the graviphoton field strength. This assumption does not hold necessarily
in the ``full'' theory. To justify this assumption and/or to compute
additional gravitational corrections one needs the supersymmetry
transformation laws in the presence of the Weyl-mulitplet. To our
knowledge these are unknown up to now. Therefore our approach bases on
perturbation theory in $W^2$. In the following we will not stress this
further, but the reader should keep in mind that, following
\cite{910}, our results hold, strictly speaking, only on the horizon without
additional corrections.}
a $1/r^2$ dependence of $T^-_{\mu\nu}$,
then it follows that the only impact of the higher order corrections
can be a change in the ``effective charge'' (= renormalization).
Thus, we use the ``renormalized'' $T^-_{\mu\nu}$, where
the higher order corrections have been already taken into account.
Moreover, in general we have for $W^2$ on the horizon
\begin{eqnarray}
  W^2_{| {\rm hor}} &=& \frac{x+iy}{r^2}.
\end{eqnarray}
Here the functions ($x,y$) depend on the charges and
represent the back reaction of the non-trivial
$W^2$-background on the black hole solution.
\subsection{Example: The heterotic S-T-U model on $T^6$}
The ``generalized prepotential'', including
higher-order curvature corrections in terms of the
Weyl-multiplet \cite{945} , reads
\begin{eqnarray}
 F(X,W^2) &=& -i \ \sum_{g=0}^{\infty} \ (X^0)^{2-2g}
 {\cal F}^{(g)}(S,T,U) \ W^{2g}
\end{eqnarray}
with classical prepotential ${\cal F}^{(0)} = -STU$
and special coordinates $S,T,U = -i z^{1,2,3}$.
The full S-duality invariant gravitational coupling in $N=4$ string theory
is given by \cite{950}
\begin{eqnarray}
\frac{1}{16 \pi} \ {\rm Re} \
\int  \frac{1}{2 \pi i} \log \eta^{24} (i S)
      {\rm tr} (R -i * R)^2
\end{eqnarray}
Using the instanton-expansion in
$q_S=e^{-2 \pi S}$
\begin{eqnarray}
 \log \eta^{24} (i S) &=& -2\pi S
                                - 24
                                \left [
                                  q_S + \frac{3}{2} q_S^2
                                  + \frac{4}{3} q_S^3 + \cdots
                                \right ]
\end{eqnarray}
one obtains the S-duality invariant form of the gravitational
coupling in the weak coupling regime
\cite{950,960}.
It follows that the corresponding higher order gravitational couplings
of the effective action are encoded in the gravitational
coupling
\begin{eqnarray}
\label{f1}
  {\cal F}^{(1)}(S) &=& - \frac{a}{2 \pi} \log \eta^{24} (i S).
\end{eqnarray}
Here we take, as usual, $a=24$.
The gravitational coupling function represents an infinite sum of
gravitational instanton effects and can be associated with
Euclidean fivebranes wrapped on $T^6$
\cite{950}. These
fivebranes are the neutral fivebranes of heterotic string theory
or equivalently the zero size fivebranes in M-theory
\cite{970,980}.
\\
The periods corresponding to the gravitational coupling function are given by
\begin{eqnarray}
 F_{0}  &=&
- i X^{0} STU
+ i  (X^{0})^{-1} W^2  S  {\cal F}_S^{(1)}
\nonumber\\
 F_{1}  &=& X^{0}  TU -  {\cal F}_S^{(1)}  (X^{0})^{-1} W^2 ,
\nonumber\\
 F_{2}  &=& X^{0} SU,
\nonumber\\
 F_{3}  &=& X^{0} ST.
\end{eqnarray}
If we restrict ourselves to axion-free configurations
$\bar X^0 X^A + \bar X^A X^0 = 0$
the stabilisation equations yield the following set
of algebraic equations.
\begin{eqnarray}
 H_{0}  &=&
 (X^{0} + \bar X^0) STU
-
\left (\frac{W^2}{X^0} + \frac{\bar W^2}{ \bar X^0} \right )
S  {\cal F}_S^{(1)}
\nonumber\\
 H_{1}  &=& i (X^{0}-\bar X^0) TU
          - i {\cal F}_S^{(1)}
               \left (
                \frac{W^2}{X^0} - \frac{\bar W^2}{ \bar X^0}
               \right ),
\nonumber\\
 H_{2}  &=& i (X^{0} - \bar X^0) SU ,
\nonumber\\
 H_{3}  &=& i (X^{0} - \bar X^0) ST .
\end{eqnarray}
Note that the metric-function $e^{-2U}$  can be expanded ``formally''
in $W^2$, i.e.
$ e^{-2U} = e^{-2U_0} + e^{-2U_1}$
with
\begin{eqnarray}
e^{-2U_0} &=&
  8 \ |X^{0}|^{2} STU,
\nonumber\\
e^{-2U_1} &=&
-2 \ \left (
       \frac{W^2}{X^0} \bar X^0 + \frac{\bar W^2}{ \bar X^0} X^0
     \right ) \ S  {\cal F}_S^{(1)}
\end{eqnarray}
This is an implicit expansion, since the moduli
still depend on the harmonic functions and the Weyl-multiplet.
In order to find the explicit expansion of the metric function
in terms of the Weyl-multiplet it is necessary to solve the
stabilisation equations.
\subsubsection{Macroscopic entropy}
Now we will
consider as an example axion-free configurations with
$X^0 - \bar X^0 = 0$ restricting ourselves
to the leading correction in $W^2$, only.
Moreover, in order to take the back reaction into account
we keep $W^2$ to be complex and introduce
$w_\pm = W^2 \pm \bar W^2$.
For this particular configuration we obtain
\begin{eqnarray}
 S,T,U &=& - \frac{1}{2} \frac{\tilde H^{1,2,3}}{X^0}, \ \ \
 \tilde H^0 = H_{2,3}=0.
\end{eqnarray}
In addition one finds an algebraic equation
to eliminate the first derivative of the gravitational coupling function
\begin{eqnarray}
 {\cal F}^{(1)}_S &=& i \ \frac{X^0 H_1}{w_-}
\end{eqnarray}
Moreover one obtains a quadratic equation in $X^0$
\begin{eqnarray}
 (X^0)^2 - \frac{i}{2} \ X^0 \ \frac{w_+}{w_-} \frac{H_1 \tilde H^1}{H_0}
  + \frac{1}{4} \frac{\tilde H^1 \tilde H^2 \tilde H^3}{H_0} &=& 0
\end{eqnarray}
with solution
\begin{eqnarray}
 X^0 &=& \frac{i}{4} \ \frac{w_+}{w_-} \ \frac{H_1 \tilde H^1}{H_0}
\pm \sqrt{- \frac{1}{4} \frac{\tilde H^1 \tilde H^2 \tilde H^3}{H_0}
+ \left (
  \frac{i}{4} \ \frac{w_+}{w_-} \ \frac{H_1 \tilde H^1}{H_0}
  \right )^2
}
\end{eqnarray}
The corresponding solution to order $W^2$ reads
\begin{eqnarray}
 X^0 &=&  - \frac{1}{2}
          \sqrt{- \frac{\tilde H^1 \tilde H^2 \tilde H^3}{H_0}}
\ \left (
  1 + \Delta
  \right ),
\nonumber\\
\Delta &=& - \frac{i}{2} \ \frac{w_+}{w_-} \
             \sqrt{- \frac{(H_1 \tilde H^1)^2}
                  {H_0 \tilde H^1 \tilde H^2 \tilde H^3}}
             + {\cal O}(W^4)
\end{eqnarray}
If one takes $H_0 \equiv -(h_0 + \frac{q_0}{r})$
it follows for
the fixed values of the moduli on the horizon to order
$W^2$
\begin{eqnarray}
 (S,T,U)_{| {\rm hor}}  &=&
 \sqrt{\frac{ (q_0 p^{1,2,3})^2}{ q_0 D}}
 \ \left ( 1 - \delta \right ).
\end{eqnarray}
with
\begin{eqnarray}
\Delta_{| {\rm hor}} &\equiv& \delta \ = \
- \frac{1}{2} \ \frac{x}{y} \ \sqrt{\frac{(q_1 p^1)^2}{q_0 D}},
\hspace{1cm}
D  \ \equiv \ p^{1}p^{2}p^{3}.
\end{eqnarray}
Note that the moduli obtain corrections of order
${\cal O}(1/\sqrt{k})$ with $k=p^2p^3$.
Straightforward calculation yields the result that
the black hole entropy does not receive explicit corrections of order
$W^2$ and is, therefore, independent of $q_1$
\begin{eqnarray}
 S_{BH}  &=& 2 \pi \ \sqrt{q_{0} D }.
\end{eqnarray}
Another example with the same result for the entropy has been given
in \cite{910}.
It follows that the corrections to the black hole entropy are
only implicit, i.e. the charges are ``renormalized''.
\subsubsection{Microscopic entropy}
In order to obtain the corresponding microscopic entropy we
follow the general concept of a ``renormalized'' charge
$q_0$ in contrast to the unrenormalized bare charge $q_0^{(0)}$ valid
for classical prepotential $F^{(0)}$. The magnetic charges remain
unrenormalized if one includes higher order curvature corrections.
Note that we will ``match'' the charge $q_0$ to the
microscopic entropy coming from CFTs \cite{412}.
The renormalized charge reads in general
\begin{eqnarray}
 q_0 &=& b_0 + q_0^{(0)} + b(q,p) q_0^{(0)},
\end{eqnarray}
where $b_0$ parametrizes a constant shift
and $b(q,p)$ is in general an unknown function depending
on the magnetic charges and $q_0^{(0)}$.
Note that this ansatz is justified by a general Taylor expansion.
However, we will consider
$b(q,p)=b(p)$ in the following. In the context of
low energy effective actions with $N=2$ supersymmetry, as
discussed in \cite{910},
this restriction is consistent
in the large
$q_0^{(0)}$-limit for the following reason: For
$q_0^{(0)}$ we have $b_0 \rightarrow 0$ and using the
MSW  formula \cite{900}
\begin{eqnarray}
 q_0 D &=& \frac{1}{6} q_0^{(0)} (6D + c_{2A} p^A)
\end{eqnarray}
one obtains $b(p) = c_{2A} p^A/6D$.
Thus we find the result of \cite{910}
for the renormalization of $q_0^{(0)}$:
\begin{eqnarray}
 q_0  &=& q_0^{(0)}
\left (
1 + \frac{c_{2A} p^A}{6D}
\right ).
\end{eqnarray}
Assuming now, that the function $b$ is in general independent of
the electric charge, one can find the proposed renormalization
of $q_0^{(0)}$ coming from CFTs \cite{412}
as follows:
\begin{eqnarray}
 q_0 D &=& \frac{1}{6} \ N \ c_{tot}
\end{eqnarray}
with $N = 1+q_0^{(0)} p^1$.
Straightforward calculation yields
\begin{eqnarray}
 b_0 &=& \frac{c_{tot}}{6k p^1},
\hspace{1cm}
 b(p) \ = \
\frac{c_{tot}}{6k} - 1.
\end{eqnarray}
Finally, in this setup one obtains the following renormalization of
$q_0^{(0)}$ including higher-order curvature corrections
\begin{eqnarray}
 q_0  &=& \left ( q_0^{(0)} + \frac{1}{p^1} \right )
          \ \frac{c_{tot}}{6k}.
\end{eqnarray}
Note that the precise values of ($\beta,\gamma$) do not play any role
in this setup.
\subsubsection{The gravitational instanton phase}
Now we will consider the particular weak coupling
regime including instanton corrections of order ${\cal O}(q_S)$
with $w_-=0$. In order to separate the instanton correction
we consider the limit $H_0 \rightarrow 0$.
This configuration represents a special point in moduli space,
which shows that gravitational instantons and/or
higher-order curvature corrections
can yield logarithmic subleading contributions to black hole
entropies. Thus, in this subsection we discuss
something new, i.e. the following discussion is not strongly related
to the rest of this article.
\\
Solving the stabilisation equations one obtains
(without expansion to order $W^2$)
\begin{eqnarray}
 X^0 &=&  \pi \
\frac{ \tilde H^1}
     {\log \left ( \frac{1- \Delta}{24} \right ) },
\hspace{2cm}
\Delta \ = \ \frac{\tilde H^2 \tilde H^3}{2a w_+ }.
\end{eqnarray}
The solution for the moduli $S,T,U$ follows straightforward
and the entropy reads
\begin{eqnarray}
 S_{BH} &=& \frac{k}{2} \  \log \left | \
  \frac{24}{1- \delta(k)} \
\right |
\end{eqnarray}
with $\Delta_{| {\rm hor}} \equiv \delta (k)$.
In this particular limit in moduli space
the classical entropy vanishes and the
black hole enters  a ``gravitational instanton phase''.
The corresponding entropy contains only logarithmic
subleading contributions and is independent of the oscillator number
$N$ provided $w_+$ is independent of $N$.
It follows that the corresponding degeneracy of states $d$ of the
underlying quantum theory is given by polynomial subleading
contributions, only:
\begin{eqnarray}
 d(k) \ = \ e^{S_{BH}} &=&
\left |
 \frac{24}{1 - \delta(k)}
\right |^{k/2}
\end{eqnarray}
Moreover, in the classical limit the degeneracy of
states vanishes
\begin{eqnarray}
 \lim_{k \rightarrow \infty} d(k)  &=& 0.
\end{eqnarray}
This shows that the inclusion of fivebrane instantons
yields non-perturbative gravitational contributions
to the black hole entropy. Here we included for convenience
only the first-order non-perturbative instanton correction, but in
general the complete S-duality invariant contributions have to
be taken into account.
\\
This result suggests the following
geometrical picture: The classical black hole can be described
by the classical background and the corresponding
action. In the limit where the
classical black hole area shrinks to zero higher order
curvature corrections must be taken into account, too. In this
limit the black hole enters a gravitational instanton phase
and the corresponding area is much smaller than the classical
one but non-vanishing in general,
unless the black hole itself ``disappears'', i.e.
the level $k$ becomes zero.
It follows that the black hole is extremely stable, i.e.
higher-order non-perturbative instanton corrections
can stabilize a black hole solution.


\section{Conclusions}

In this article we discussed the CFT for $AdS_3$ gravity with a
spatial annulus geometry, which appears naturally if a BTZ black hole
is excited.  We paid special attention to an exact treatment, i.e.\
we did not assume a large $N$ and/or $\alpha' \rightarrow 0$
expansion. As a concrete model we considered the near-horizon geometry
of a non-extreme 4-charge configuration comprising a fundamental
string with wave modes, a 5-brane and Taub-NUT soliton. The spherical
part is given by an $S_3/Z(m)$ geometry which is described by a CFT
given by an $SU(2)/U(1)$-WZW model. In the $AdS_3$ part the momentum
modes excite a BTZ black hole and therefore the spatial geometry
has two boundaries: The asymptotic one and the horizon of the black
hole.
A careful treatment shows that one finds
two different CFTs at these two boundaries:
At the asymptotic boundary it is a Liouville model and at the horizon
a 2-d black hole. Both CFT's can be expressed by an $SL(2, {\bf R})$
coset, but the $U(1)$ that has to be gauged differs; on the horizon it
is spatial whereas at the infinity it is a lightcone direction. The
BTZ black hole can therefore be seen as a solution interpolating
between these two cosets on the boundaries. Thus, there are
two types of boundary states: States
living on the asymptotic boundary and on
the BTZ-horizon. Both types of states
contribute to the total central charge, which is
\begin{eqnarray}
 c_{tot} = c_{SU} -1 \, + c_{2d-BH} + c_{L}
         = \Big({3k \over k+2 } -1 \Big) +  \Big({3k \over k-2} -1\Big) +
\Big({3k \over k-2} -2 + 6k\Big).
\end{eqnarray}
It follows that in the classical limit $\alpha' \rightarrow 0$ or
$k \rightarrow \infty$ only the Liouville part on the asymptotic
boundary contributes and gives the well-known result $c_{class}
=6k$.
\\
An interesting observation is that in the supersymmetric case the $k$
dependence of the $SU(2)$ model cancels with the $k$ dependence of the 2-d
black hole. The central charge becomes $c = 6(k + {1 \over k}) +
const.$, which is invariant under $k \rightarrow 1/k$. As discussed
below eq. (\ref{802}), at the self-dual point a ``massless tachyon''
appears and it is interesting to note, that the $1/k$ term produces an
energy gap as discussed in \cite{981} (note that, when applied
to black holes, the central charge is directly related to the
minimal mass).
\\
Our 2-boundary setup may also imply an interesting worldvolume
interpretation. From the worldvolume point of view, the
radial coordinate of the $AdS$ space sets the energy scale
and the boundary CFTs of the $AdS$ space are expected to be
dual to a worldvolume CFT at the renormalization group fixpoint
(vanishing $\beta$-functions). Therefore every boundary CFT corresponds
to a different fixpoint in the worldvolume theory and moving
from one fixpoint to another corresponds to going from one
$AdS$ boundary to another.

In the second part we tested the proposal of a statistical
(microscopic)
interpretation of the (macroscopic)
Bekenstein-Hawking entropy coming from
CFTs. In particular we used the $AdS$/CFT correspondence and
showed by an independent field theoretical calculation that
the results are consistent. However, strictly speaking
the results presented do not ``prove'' anything, since
our approach, given in \cite{910}, is rather limited.
On the other hand, since all the results are consistent,
we believe that our approach to obtain a statistical
interpretation of the Bekenstein-Hawking entropy,
including all $\alpha^\prime$ corrections,
represents a good perspective for future investigations.
\\
Apart from these results we have presented a special limit in moduli
space, where the classical black hole entropy vanishes.
However, including non-perturbative gravitational instantons
the Bekenstein-Hawking is non-vanishing, depends only
on the level $k$ and is logarithmic.
This result shows that
gravitational instantons can stabilize a black hole solution and that
logarithmic subleading black entropies can in principle arise
in models with $N>2$ supersymmetry, too.


\bigskip \bigskip

{\bf Acknowledgments}

We would like to thank S. F\"orste, H.\ Dorn, H.-J.\ Otto,
M. Walton and M. Gaberdiel for discussions.

\medskip

\newcommand{\NP}[3]{{ Nucl. Phys.} {\bf #1} {(19#2)} {#3}}
\newcommand{\PL}[3]{{ Phys. Lett.} {\bf #1} {(19#2)} {#3}}
\newcommand{\PRL}[3]{{ Phys. Rev. Lett.} {\bf #1} {(19#2)} {#3}}
\newcommand{\PR}[3]{{ Phys. Rev.} {\bf #1} {(19#2)} {#3}}
\newcommand{\IJ}[3]{{ Int. Jour. Mod. Phys.} {\bf #1} {(19#2)}{#3}}
\newcommand{\CMP}[3]{{ Comm. Math. Phys.} {\bf #1} {(19#2)} {#3}}
\newcommand{\CQG}[3]{{ Class. Quant. Grav.} {\bf #1} {(19#2)} {#3}}
\newcommand{\PRp} [3]{{ Phys. Rep.} {\bf #1} {(19#2)} {#3}}
\newcommand{\MPL}[3]{{Mod. Phys. Lett.} {\bf #1} {(19#2)} {#3}}

\end{document}